\documentclass[preprint,12pt]{aastex}

\begin{document}

\newcommand\tna{\,\tablenotemark{a}}
\newcommand\tnb{\,\tablenotemark{b}}

\title{The Mid-Infrared Instrument for the James Webb Space Telescope, VII: The MIRI Detectors}

\author{G. H. Rieke\altaffilmark{1}, M. E. Ressler\altaffilmark{2},  Jane E. Morrison\altaffilmark{1}, L. Bergeron\altaffilmark{3}, Patrice Bouchet\altaffilmark{4}, Macarena Garc\'ia-Mar\'in\altaffilmark{5}, T. P. Greene\altaffilmark{6},
 M. W. Regan\altaffilmark{3}, K. G. Sukhatme\altaffilmark{2}, \& Helen Walker\altaffilmark{7}}

\altaffiltext{1}{ Steward Observatory, 933 N. Cherry Ave, University of Arizona, Tucson, AZ 85721, USA}
\altaffiltext{2}{Jet Propulsion Laboratory, California Institute of Technology, 4800 Oak Grove Dr. Pasadena, CA 91109, USA}
\altaffiltext{3}{Space Telescope Science Institute, 3700 San Martin Drive, Baltimore, MD 21218, USA}
\altaffiltext{4}{Laboratoire AIM Paris-Saclay, CEA-IRFU/SAp, CNRS, Universit\'e Paris Diderot, F-91191 Gif-sur-Yvette, France}
\altaffiltext{5}{I. Physikalisches Institut, Universit\"at zu K\"oln, Z\"ulpicher Str. 77,  50937 K\"oln, Germany }
\altaffiltext{6}{NASA Ames Research Center, M.S. 245-6, Moffett Field, CA 94035, USA}
\altaffiltext{7}{RALSpace, STFC, Rutherford Appleton Lab., Harwell, Oxford, Didcot OX11 0QX, UK}

\begin{abstract}

The MIRI Si:As IBC detector arrays extend the heritage technology from the {\it Spitzer} IRAC arrays to a 1024 x 1024
pixel format. We provide a short discussion of the principles of operation, design, and performance of the individual MIRI detectors, 
in support of a description of their operation in arrays provided in an accompanying paper (\citet{ressler2014}). We then describe
modeling of their response. We find that electron diffusion is an important component of their
performance, although it was omitted in previous models. Our new model will let us optimize the bias voltage while
avoiding avalanche gain. It also predicts the fraction of the IR-active layer that is depleted (and thus contributes to the quantum efficiency) as
signal is accumulated on the array amplifier. Another set of models accurately predicts the nonlinearity of
the detector-amplifier unit and has guided determination of the corrections for nonlinearity. Finally, we discuss how
diffraction at the interpixel gaps and total internal reflection can produce the extended cross-like
artifacts around images with these arrays at short wavelengths, $\sim$ 5 $\mu$m. The modeling 
of the behavior of these devices is helping optimize how we operate them and also providing
inputs to the development of the data pipeline. 

\end{abstract}


\section{Introduction}

The detectors of choice for the 5 - 28 $\mu $m range are arsenic-doped silicon impurity 
band conduction (Si:As IBC) devices. They have extensive space flight 
heritage, for example, arrays of these devices were used 
in all three Spitzer instruments (IRAC: \citet{hora2004a}; IRS: \citet{vancleve1995},  
\citet{houck2004}; and MIPS: \citet{gordon2004}), in WISE \citep{mainzer2008}, 
in MSX \citep{mill1994} and in Akari \citep{onaka2007}. The focal planes on these missions have 
demonstrated high detective quantum efficiencies, low dark current 
and relative freedom from other spurious signals, excellent photometric performance, and 
resistance to the effects of cosmic radiation. Similar detector arrays were selected 
for the Mid-Infrared Instrument (MIRI) on JWST. They have been constructed 
at Raytheon Vision Systems, specifically based on the experience at this 
supplier with the detector arrays for IRAC \citep{wu1997}. The readout integrated circuit (ROIC), electronic data chain, 
mechanical arrangement, and operation of the MIRI arrays are described in \citet{ressler2014}, hereafter Paper VIII; 
while Gordon et al. (2014; Paper X) describe plans to calibrate the
detector signals, including removal of the artifacts discussed below.  The current 
paper discusses the individual detectors themselves.After an introduction to the basic detector properties
in Section 2, in Section 3 we describe work to understand the detector performance from first
principles as a way to help optimize the instrument performance and to support the
development of the data reduction software. 

\section{Detector design and performance}

\subsection{Detector design}

\begin{figure}[htbp]
\centerline{\includegraphics[width=5.99in]{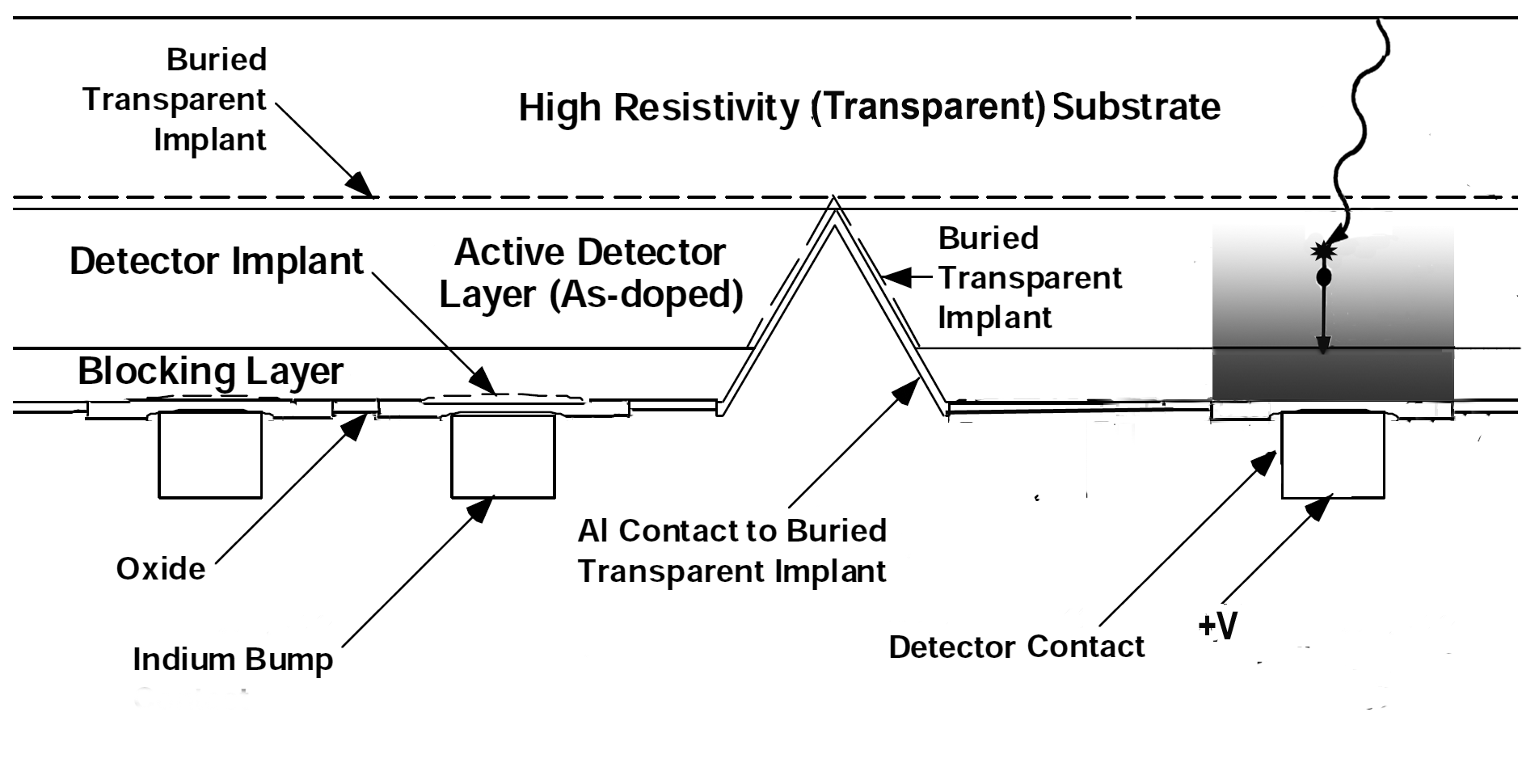}}
\caption{Basic operation of a Si:As IBC detector array. The left side shows the physical arrangement of the detectors. They are manufactured on a lightly doped (high resistivity) transparent substrate. A 25 - 35 $\mu$m thick layer is relatively heavily 
doped with arsenic and absorbs the incident photons. An electric field is maintained across this layer by putting a voltage between
the detector contact and the buried transparent common contact; connection to the buried contact is made through the V-etch (to one side of the detector array). The field is shown schematically by the gray-scale on the right pixel. An 
 intrinsic layer blocks thermally generated free charge carriers in the arsenic impurity level from escaping from the IR-active layer and reaching the detector contact  
(because there is no corresponding energy level in intrinsic material), but allows the
photoelectrons passage since they have been elevated into the conduction band.  
A photon is shown generating a photoelectron just at the edge 
of the depletion region where there is sufficient electric field to draw the free electron to the appropriate detector contact. 
When the array is hybridized to the readout wafer, contact to an output amplifier is made through an individual indium bump for each pixel. Figure based on that in Love et al. (2005).} 
\label{fig:fig1}
\end{figure}

\begin{figure}[htbp]
\centerline{\includegraphics[width=4.0in,height=3.0in]{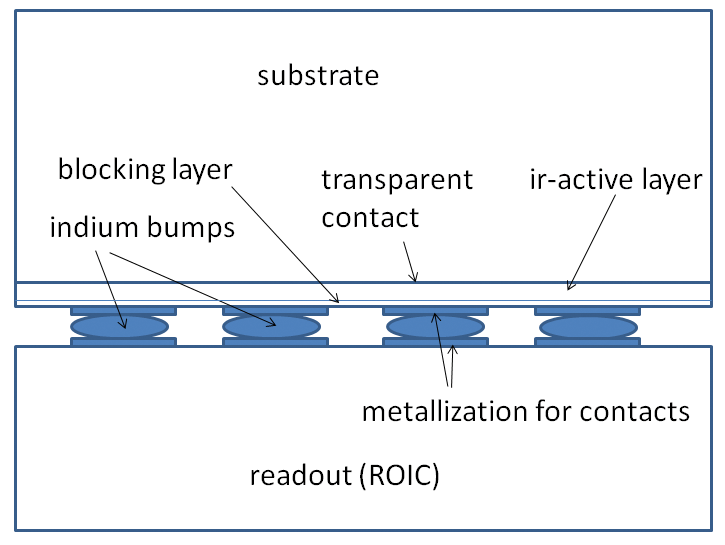}}
\label{fig11}
\caption{Illustration of the geometry of an arsenic-doped silicon detector array.  The light travels through ~500 $\mu$m of pure silicon (which may be anti-reflection coated) and then a transparent contact into a thin active layer (25 $\mu$m for IRAC, 35 $\mu$m for MIRI baseline material) of arsenic-doped silicon, followed by an even thinner (4 $\mu$m for MIRI) ‘blocking layer’ of pure silicon.  Indium bump bonds, one per pixel, join evaporated contacts on the detectors and readout amplifiers to convey the photo-electrons generated in the active layer to the input of the amplifiers.}
\label{fig:fig2}
\end{figure}

The structure of the detector layers (alternately referred to as Blocked
Impurity Band---BIB---or Impurity Band Conductor---IBC device, depending on the
vendor) is shown in Figure 1. The pixels are manufactured on a wafer that provides 
a transparent silicon substrate, on which is produced a buried transparent 
contact and then an infrared-active detector layer of 
thickness 25 - 35$\mu $m, relatively heavily doped (with arsenic for the MIRI devices) 
to produce an impurity energy band. 
This layer is followed by an intrinsic blocking layer about 3 - 4$\mu $m 
thick, with the second contact on it. This contact defines the pixel and
is indium-bump-bonded to the input of the ROIC amplifier (see Figure \ref{fig:fig2}). The bias voltage on the pixel 
is established between its individual contact and the buried common one whose bias is maintained 
through a V-shaped etched trough, aluminum-coated 
to make it conductive, and placed to one side of the array (Figure \ref{fig:fig1}), although 
for illustration the figure shows it in the middle. 

Impurity band 
conductivity in the IR-active layer would produce unacceptable 
dark current if the band had electrical access to both detector 
contacts. However, the impurity band terminates at the intrinsic blocking layer, so 
leakage can only occur through thermal excitation up into the conduction 
band (this discussion is for an n-type dopant, e.g. arsenic). So long as the detectors 
are sufficiently cold (e.g., below 6.7K), this form of thermal current is very small.  
Because the intrinsic layer blocks the dark current, the doping in the 
infrared-active layer can be two orders of magnitude greater than in a bulk 
photoconductor\footnote{That is, a detector consisting of an undifferentiated 
layer of photoconductive material placed between two electrical contacts.}. 
The highly doped, relatively thin IR-active layer overcomes the issues 
with bulk detectors of slow adjustment toward electrical equilibrium and of 
large cross section for ionizing radiation (for more discussion of these detectors, 
see, e.g., \citet{petroff1985}; \citet{rieke2002}) .

To collect the photo-generated charge carriers, the electrical conductivity of the 
IR-active layer must be suppressed to allow a field to be generated across it. 
When a positive bias is applied across the 
blocking layer to the buried contact, negative carriers are collected at the 
interface between the intrinsic blocking and IR-active layers and positive ones 
are driven away from this interface. Thus, a high-resistance region depleted 
of free charge carriers is produced in the IR-active material near the 
interface. However, beyond the depletion region there is virtually no field; 
this situation is shown schematically to the right in Figure 1 where the field is 
indicated by gray-shading and it dies away before reaching completely across the IR-active layer.  
Photoelectrons are produced throughout the IR-active layer; to gather them efficiently requires that it be fully 
depleted\footnote{Diffusion lengths of a few $\mu$m also allow some charge collection from 
regions adjacent to those that are depleted.}. The minority, p-type impurities attach electrons and 
maintain a negative space charge in this region that tends to cancel the 
effect of the positive bias. The expression for the width of the depletion region
 shows this effect:

~~~~~~~~~~~~~~~$w=\left[ \frac{2\thinspace {}\kappa_{0}\thinspace \varepsilon 
_{0}}{q\thinspace N_{A}}\left| V_{bias} \right|+t_{B}^{2} 
\right]^{1/2}-t_{B}$   ~~~~~~~~~~~~~~~~~~~~~~ (1)

\noindent
where $N_{A}$ is the p-type impurity concentration, $q$ is the electronic 
charge, $\kappa_{0}$ is the dielectric constant 
of silicon, $\varepsilon_{0}$ is the permittivity of free space, \textbar 
$V_{bias}$\textbar~ is the bias voltage across the detector contacts, and 
$t_{B}$ is the thickness of the blocking layer \citep{rieke2002}. Equation (1) implies that large values 
of $N_A$ can be overcome with large bias voltages. However, if the voltage across the 
detector is made too large, the field near the blocking layer can become 
strong enough that the photoelectrons gain sufficient energy to free 
additional electrons, creating avalanche gain. This gain is undesirable 
because it is generated by a stochastic process and adds noise to the signal. Thus, there 
is an imperative in keeping the minority impurity concentration low to make 
$w$ large at a modest level of $V_{bias}$. Complete depletion in the MIRI arrays 
without significant gain requires that the level of the minority impurities 
be kept below about 2 X 10$^{\mathrm{12\thinspace 
}}$cm$^{\mathrm{-3}}$. That is, the impurity atoms must be at or below 1 in 
10$^{\mathrm{10}}$ silicon atoms. 

In Figure 1, light enters from
the top and passes through the thick, inactive substrate layer of pure (i.e.,
high resistivity) silicon. After also passing through the transparent electrical
contact, the photon is absorbed in the active layer of
arsenic-doped silicon, as shown to the right in Figure 1. The arsenic concentration can be 
made high enough for almost complete absorption in the 10 - 25 $\mu$m range, where the
dominant losses occur at the transparent contact. However, the absorption cross section is wavelength dependent,
and short wavelength photons may pass through the active layer, reflect off
the metallized electrical contact, and be absorbed as they travel back out
the active layer (or they may escape from the initial pixel, see discussion in Section 3.3.2 below).  
There are no physical boundaries between pixels in an IBC detector array --- the
detectors are all on a single slab of silicon. Pixels are defined by
the very high electric fields induced between the transparent contact
separating the active and inactive layers and the contact at the indium
bumps. The fields are so high that a created electron-hole pair is quickly
swept out the nearest contact, so that interpixel diffusion is generally not
an issue.

The detector material for MIRI was grown at Advanced 
Technology Material, Inc. (ATMI), where precautions were taken to achieve 
high purity levels. Spreading resistance measurements suggest that the 
minority (acceptor) impurities were held to a density of 2 X 
10$^{\mathrm{12}}$ cm$^{\mathrm{-3}}$ or less \citep{love2005}. 
Two designs were implemented. The 
``baseline" arrays have an arsenic doping level of 7 X 10$^{\mathrm{17}}$ 
cm$^{\mathrm{-3\thinspace }}$ and an infrared-active layer 35 $\mu $m thick, 
while the "contingency" arrays are similar but with the IR-active 
region reduced to 30 $\mu $m in thickness and an arsenic concentration
of 5 X 10$^{\mathrm{17}}$ 
cm$^{\mathrm{-3\thinspace }}$. Both types have blocking layers 
of 4 $\mu $m thickness. The contingency material was designed to be virtually identical
to that used in the IRAC arrays and thus to have a high probability of reproducing their performance. 
The baseline design was developed for the
potential of higher quantum efficiency where the absorption
cross sections are small (e.g., at wavelengths short of 8 $\mu$m) using a thicker infrared-active layer. 
The price is a higher dark current: about 0.2 e/s at the 6.7K operating temperature, about three
times higher than for the contingency material. 

\begin{figure}[htbp]
\centerline{\includegraphics[width=3.0in,height=2.3in]{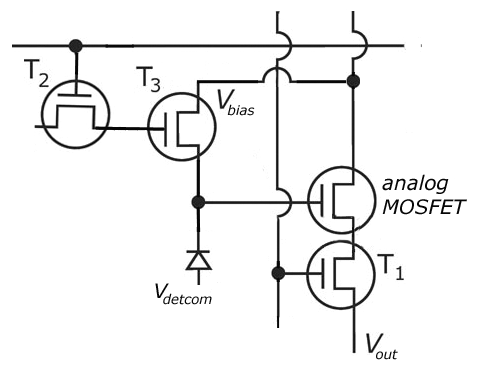}}
\label{fig2}
\caption{A representative Si:As IBC detector array unit cell. The circuit is a simple source follower, where current 
from the detector (drawn as a diode to indicate its electrical assymetry) delivers charge to the gate of the analog MOSFET amplifier. 
$V_{out}$ is proportional to the charge integrated onto the gate. 
The unit cell is addressed by T$_1$ (column) and T$_2$ (row), and T$_3$ is used to set the detector voltage
back to the initial bias level. The design is based on that of the unit cell for the IRAC arrays.} 
\label{fig:fig3}
\end{figure}

To manufacture detector arrays, the detector wafers were diced and patterned 
with indium bumps, allowing them to be hybridized onto the readouts (Figure \ref{fig:fig2}).  Each 
detector is connected to the input of a source follower amplifier as shown in Figure \ref{fig:fig3}. 
The array format is 1024 X 1024 
pixels with a pitch of 25 $\mu $m. The 
hybridized arrays are then anti-reflection (AR) 
coated with either of two single layer AR thicknesses, one optimized for 6 $\mu 
$m and the other for 16 $\mu $m. These steps complete the construction of a 
MIRI Sensor Chip Assembly (SCA). The properties of the MIRI Si:As IBC 
SCAs are described by \citet{love2004, love2005, love2006} 
and \citet{ressler2008}, and have a heritage in both the detectors and ROIC to 
the arrays developed for IRAC \citep{wu1997, hora2004a}. 

\subsection{Performance}

We introduce a few basic properties of the MIRI detectors to motivate 
the discussion of the operation of the devices in Section 3. The basic parameters are
summarized in \citet{ressler2008} and in Table 1. 

\subsubsection{Radiometric properties}

Figure \ref{fig:fig4} shows the response of the baseline detector material as a function 
of bias voltage at a wavelength near 13$\mu$m. The general characteristics are a rise in response
from zero bias to a bias of $\sim$ 1.2 V as the increasing voltage depletes more of the infrared-active layer,
followed by a leveling off for biases of 1.2 V $<$ V $<$ 2.5 V , with finally the beginning of an upturn
in the response above 2.5 V as avalanche gain begins to become significant. 
As the detector photocurrent is collected on the input of the unit cell amplifier (Figure 3),
the net voltage across the detector decreases and the response to further illumination is
reduced according to the relation illustrated in Figure 4. 
As a result, the integration ramps have a smooth departure from linearity;
just before the signals saturate, the reduction is about 10\% relative to a linear
extrapolation of the small-signal behavior. 

Figure \ref{fig:fig5}  shows the measured responsive quantum efficency (using a large process control
detector) as a function of wavelength, and the values computed from the same measurements
assuming single layer anti-reflection coatings optimized for 6 and 16$\mu$m. 

\subsubsection{Imaging properties}

The response of the detectors in an array is uniform, with pixel-to-pixel variations of order
 3{\%} rms. The best arrays have a small proportion of inoperative 
pixels (either hot or dead), of order 0.1{\%}. 

The detectors have a complex latent image behavior. 
After removing a bright source, there is an initial fast decay with a time constant of about 8 seconds. 
After this signal has died away, there are latent images at the
initial level of about 1\% of the initial image, in observations with a constant 
3-second readout cadence. The fading of these images shows multiple time constants, as is shown in Figure \ref{fig:fig6}.
Extrapolating the three exponential fits in the figure back to the time the source
was removed indicates an initial latent image of $\sim$ 1\% of the illumination, in
agreement with measurements. Therefore, the three-exponential fit is not missing 
any significant faster effect. However, it appears that the details of the fit change
according to the strength of the initial signal; it is not adequate to simply scale the
response from one signal strength to another to fit the latent image behavior.

The primary effect of the latent images is a depression of the reset level that is slowly restored to
its pre-exposure value. The result is that the fits to the integration ramps are steepened, creating an
apparent positive image that decays in strength as the depression fades. 
The latent image behavior is similar to that observed previously with similar detectors. 
For example, the primary latent decay time constant for the MIRI array of 8 seconds
is very similar to the 12 $\pm$ 5 seconds reported for the IRS/MIPS arrays (Gordon et al. 2005). 
The MIPS arrays also had multiple time constant latent decay patterns, such as long-life latent images after hard saturation \citep{gordon2004}. 
The IRAC arrays had similar behavior \citep{hora2004a}.

The imaging properties of MIRI were measured multiple times during the 
buildup of the instrument modules and then in the test of the flight model 
prior to delivery, and finally in the Integrated Science Instrument Module 
(ISIM) test post delivery. The most precise of these measurements in terms 
of the array performance were conducted at the Commissariat \`a l'\'energie 
atomique (CEA) with just the imager, and 
illuminated by a source outside the cryostat with extensive filtering to 
control the background emission (e.g., \citet{ronayette2010}). The point 
spread function (PSF) measurements utilized a micro-stepping strategy so 
that many positions of the source were recorded, on centers smaller than the 
pixel pitch of the MIRI array. These measurements were then converted to a 
high-resolution PSF image. The images at the longer wavelengths are as 
expected (Figure \ref{fig:fig7}), showing excellent imaging properties from the array. 
However, at 5.6 $\mu $m there is an additional 
cross-like imaging artifact as shown in Figure \ref{fig:fig8}.  Similar artifacts were seen
with the IRAC arrays (Figure \ref{fig:fig8}).

At all wavelengths, there is a low level of pixel-to-pixel crosstalk, which we have measured in a number of ways (\citet{finger2005}, 
\citet{regan2012}, \citet{rieke2012}), including 
autocorrelation, analysis of cosmic ray hits, and of hot pixels. All 
measurements indicate a level close to 3{\%} for the crosstalk to the four 
adjacent pixels around the one receiving signal. The dominant cause of this 
behavior is interpixel capacitance, although there may be secondary 
contributions from electron diffusion and optical effects \citep{regan2012, rieke2012}.

\section{Modeling}

\subsection{Responsivity}

The theory of silicon impurity band conduction detectors (Si:X IBC) was 
developed in a series of papers published in the 1980s (\citet{petroff1984, petroff1985, 
szmulowicz1987, szmulowicz1988}). However, those studies 
were of detectors with minority impurity concentrations of 
10$^{\mathrm{13}}$ - 10$^{\mathrm{15}}$ cm$^{\mathrm{-3}}$, majority 
impurity concentrations of about 5 X 10$^{\mathrm{17}}$ cm$^{\mathrm{-3}}$, 
and infrared-active layers 15 to 25 $\mu $m thick. The MIRI detectors 
differ significantly with regard to the first and 
third parameters. We have therefore generated a new set of detector models, 
accounting for performance aspects that were not apparent in the older 
analyses. 

\begin{figure}[htbp]
\centerline{\includegraphics[width=4.0in,height=3.0in]{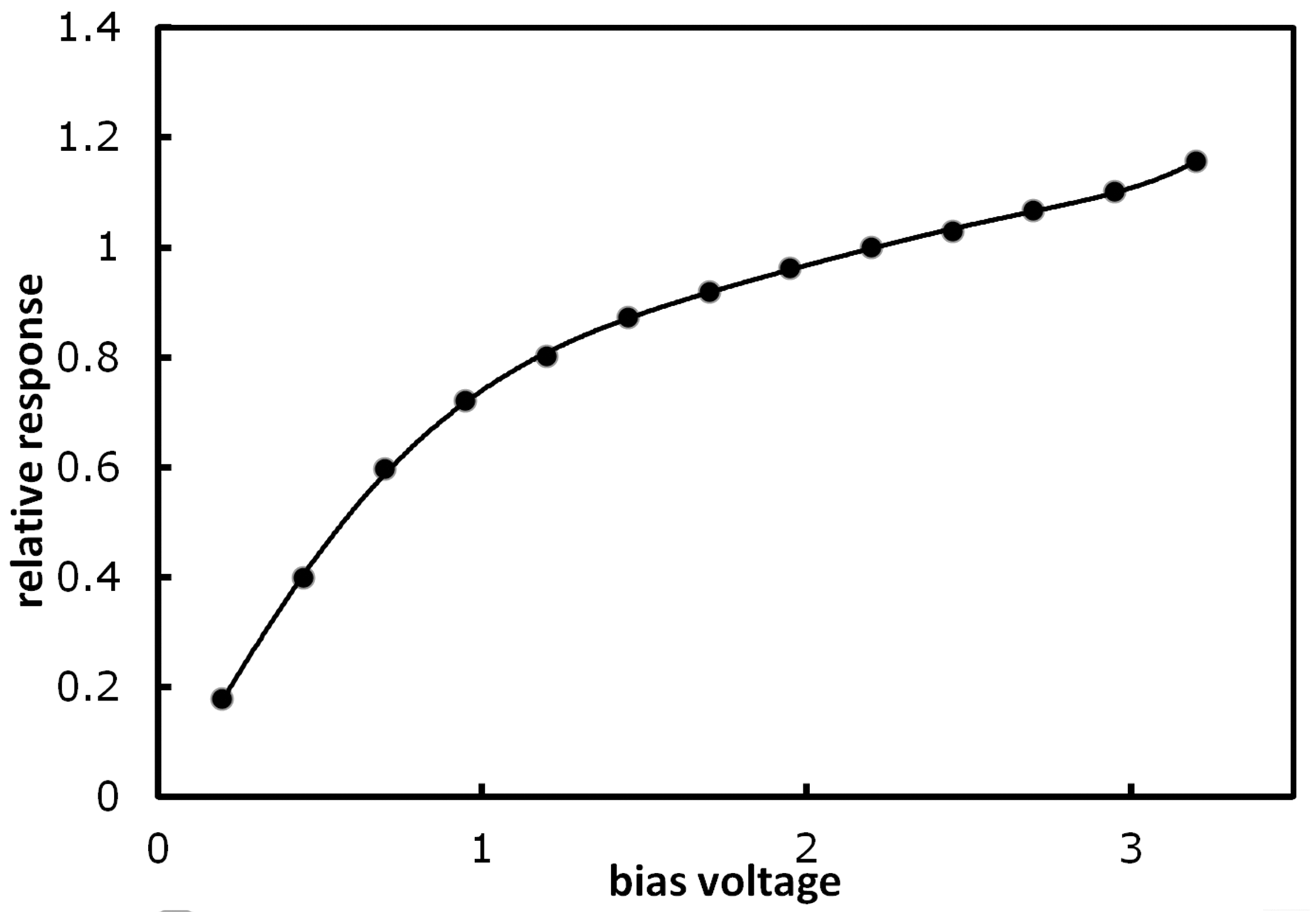}}
\label{fig4}
\caption{Relative responsivity vs. bias voltage for the baseline material \citep{ressler2008}. Good control of the minority impurity in the IR-active layer is indicated both by the knee near 1.2V which indicates that full depletion of that layer is being approached at that voltage, and the slow onset of avalanche gain above 2.5V.} 
\label{fig:fig4}
\end{figure}

\begin{figure}[htbp]
\centerline{\includegraphics[width=4.0in,height=3.0in]{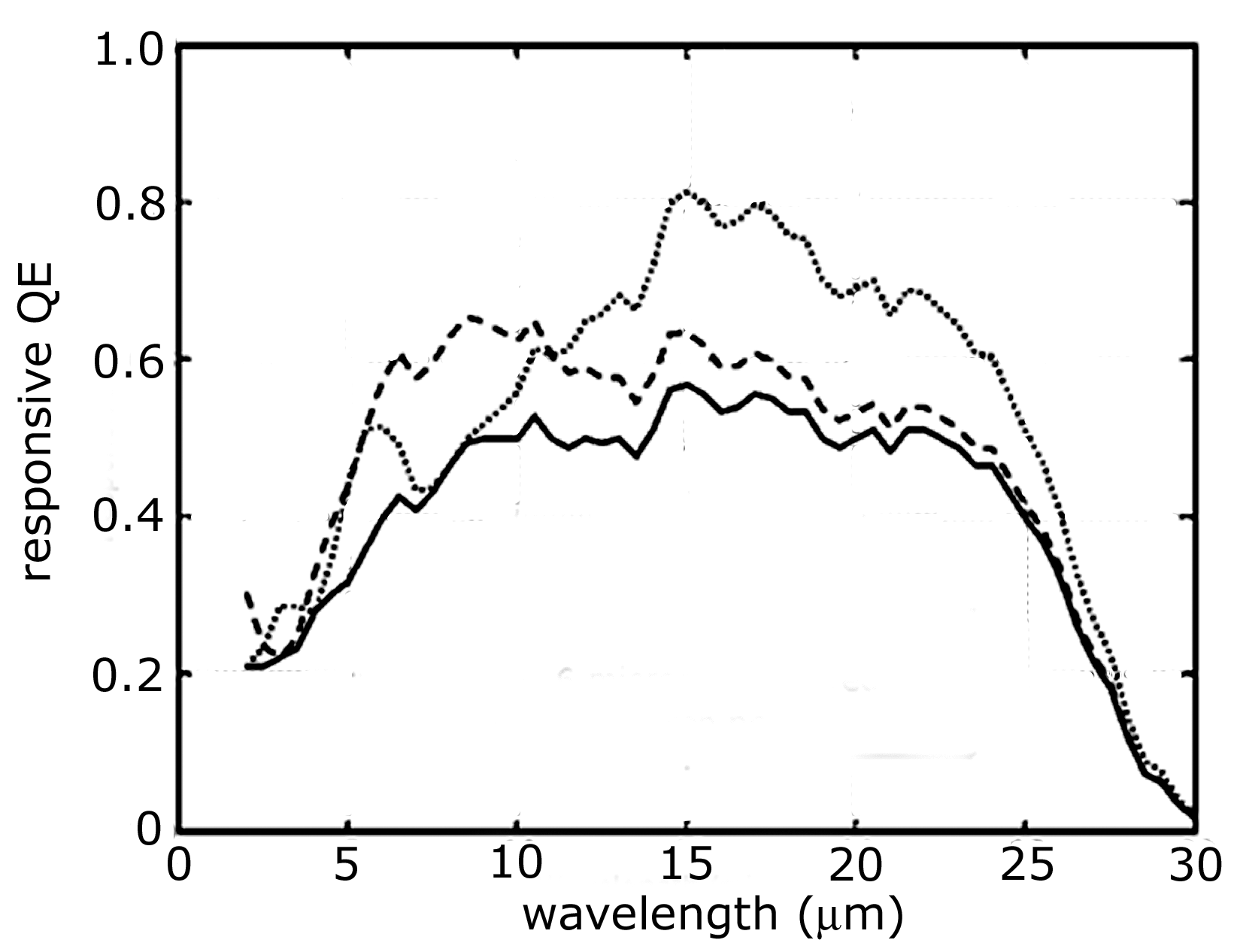}}
\label{fig6}
\caption{Measured responsive quantum efficiency of bare detector material (solid line). The dashed line is a computed result assuming the array has an antireflection coating applied optimized for 6 $\mu$m, and the dotted line is for an AR coating optimized for 16 $\mu$m.
This figure is based on one in \citet{ressler2008}. } 
\label{fig:fig5}
\end{figure}

\begin{figure}[htbp]
\centerline{\includegraphics[width=5.5in,height=4.0in]{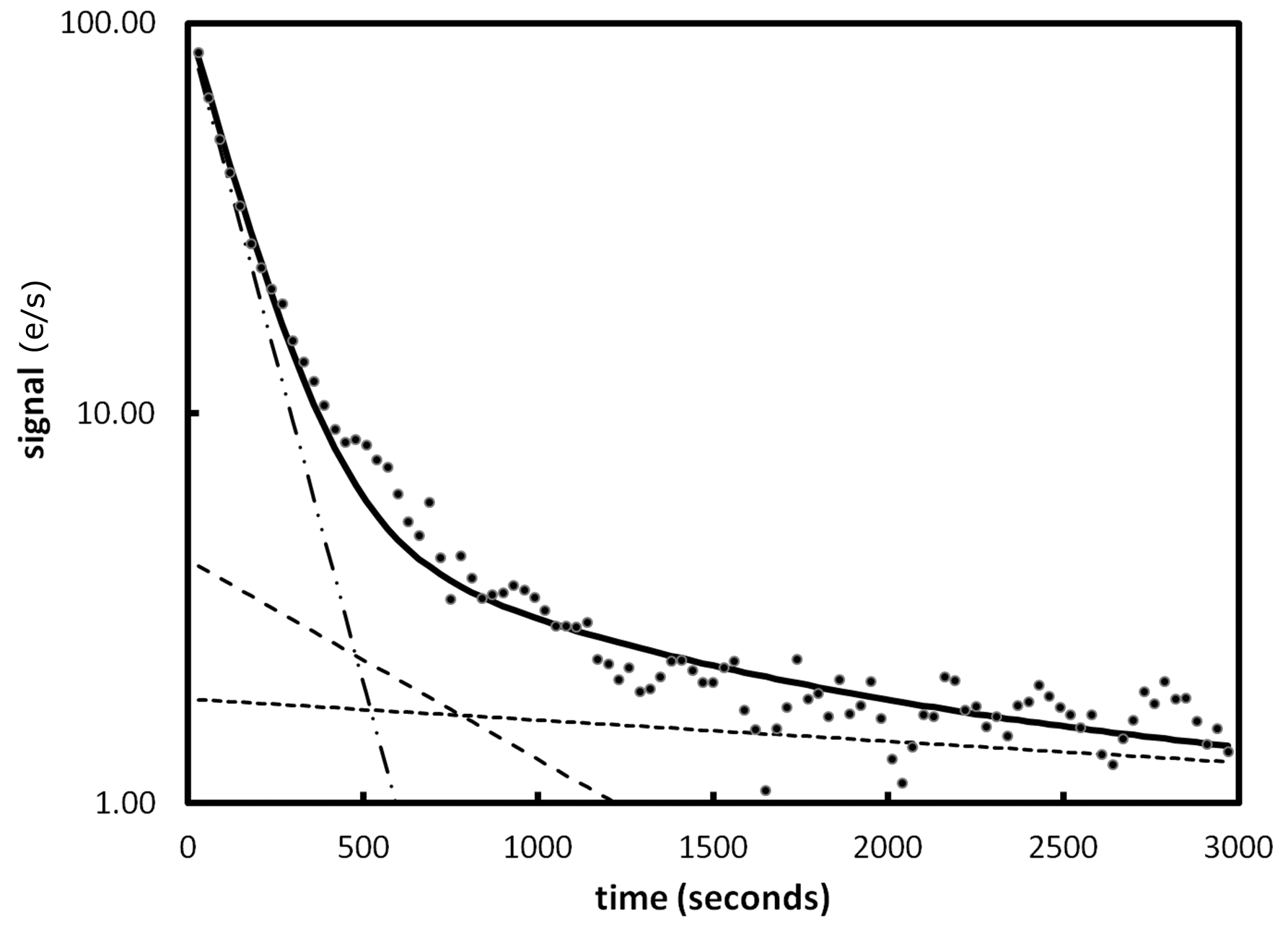}}
\label{fig6a}
\caption{Recovery from a saturating signal. The signal was about 20,000 e/s and overfilled the amplifier wells by about a factor of two (i.e., about 500,000 total electrons); there is about a 3 minute interval between removing the signal and the start of data collection, to allow placing the detector in the dark. The signal (points) has been fitted with three exponentials: dot-dash, timeconstant of 130 seconds; long dashed, time constant of 14 minutes; short dashed, time constant of 130 minutes. The
two longer time constants are very uncertain because they depend strongly on the placement of the zero signal baseline.
The solid curve shows the sum of the three fits. } 
\label{fig:fig6}
\end{figure}

\begin{figure}[htbp]
\centerline{\includegraphics[width=6.0in,height=2.0in]{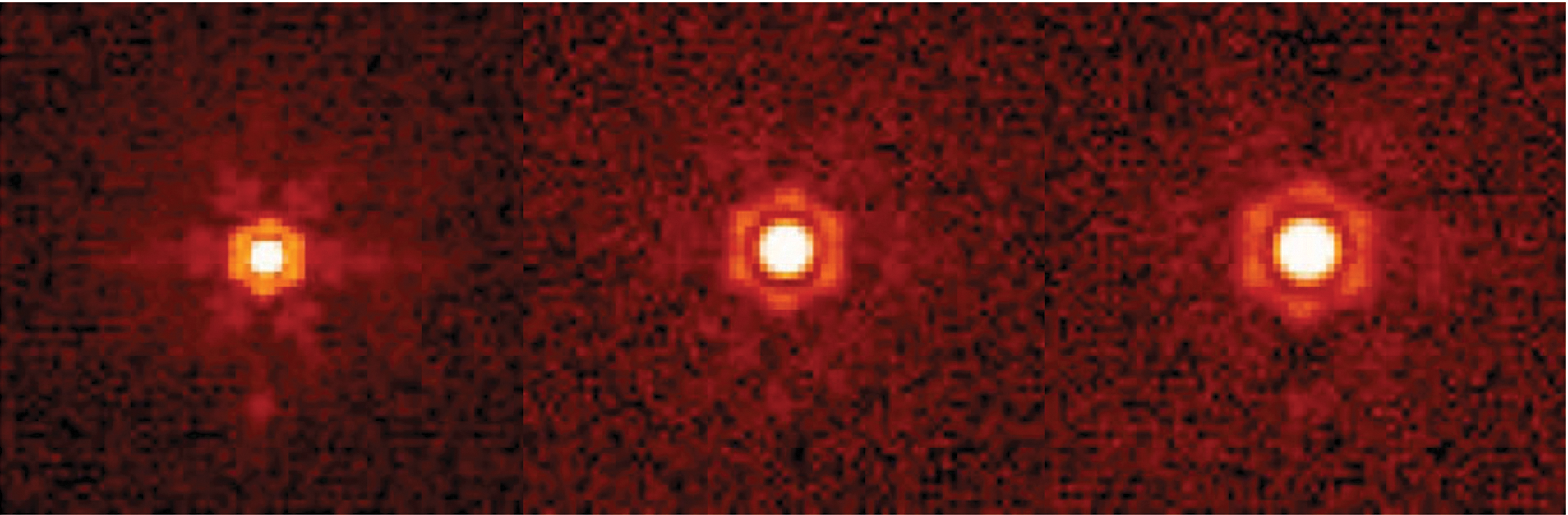}}
\label{fig7}
\caption{Measured point spread function with a simulated JWST pupil, from left to right, for 7.7 $\mu$m, 12.8 $\mu$m, and 15 $\mu$m
\citep{ronayette2010}.}
\label{fig:fig7} 
\end{figure}

\begin{figure}[htbp]
\centerline{\includegraphics[width=6.0in,height=3.0in]{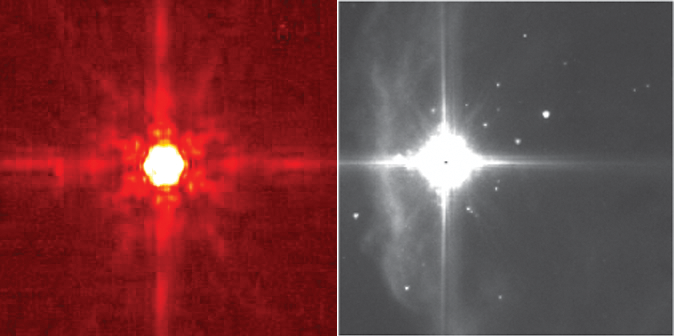}}
\label{fig8}
\caption{Left: measured point spread function for the MIRI imager at 5.6 $\mu$m \citep{ronayette2010}. Right: image with IRAC at 5.8$\mu$m of the bright 
protostar S140 \citep{pipher2004}. The cross response is an artifact resulting from the low absorption cross section for these photons in the IR-active layer.} 
\label{fig:fig8}
\end{figure}

The model for detector responsivity will be illustrated for the baseline detectors. 
The basic input data are contained in Figure \ref{fig:fig4}, and 
the parameters of the detectors are described in \citet{love2004} and \citet{love2005}. 
The detailed theory of responsivity vs. bias of IBC detectors was discussed 
in Petroff and Stapelbroek (1984, 1985). In our model, 
we used equation (39) in \citet{szmulowicz1987} to quantify the 
avalanche gain as a function of depth in the detector. The gain is largely 
controlled by the characteristic field strength, $E_{{C}}$. 
The adopted value has a strong influence on the upturn in the 
responsivity curve for biases above 2.5V in Figure \ref{fig:fig4}; we leave it as 
a free parameter in our model to allow accurate fitting of this behavior. 

Previous models have assumed that the signal-producing region of the 
detector includes only that part of the IR-active layer that is depleted of 
charge carriers by the applied electric field as in equation (1) (e.g.,\citet{petroff1985, 
szmulowicz1987}). Photoelectrons produced in the 
undepleted (neutral hereafter) region are assumed to recombine without 
contributing to the signal. This assumption is difficult to reconcile with 
the behavior of the MIRI detectors. At the wavelengths near peak absorption, 
the absorption cross section is of order 1.3 X 10$^{\mathrm{-15}}$ cm$^{\mathrm{2}}$. 
Because of the relatively large thickness of the IR-active layer and the 
high concentration of As, the absorption length is only about 10 $\mu $m. In 
our model, at 0.5V bias the neutral layer is about 18 $\mu $m thick and for 
our back-illuminated detectors the absorption occurs at the side of the 
IR-active layer opposite the depleted region, leaving a gap of about 8 $\mu 
$m between the depleted region and where the photons are absorbed. Thus, one 
would conclude that there would be very little signal at biases around 0.5V, 
in disagreement with the slow reduction in response with reduced bias in 
Figure \ref{fig:fig4}. 

The missing ingredient is diffusion of the free 
charge carriers from their places of origin. 
There is no analytic form for 
the general case of diffusion of charges within a plane-parallel detector, 
so we have used the expression for the case 
where the absorption occurs all at the entrance to an infinite 
planar detector and diffuses across a region of thickness $c$ \citep{holloway1986}:

\[
S=\frac{2b}{e^{\raise0.7ex\hbox{$c$} \!\mathord{\left/ {\vphantom {c 
L}}\right.\kern-\nulldelimiterspace}\!\lower0.7ex\hbox{$L$}}+e^{\raise0.7ex\hbox{${-c}$} 
\!\mathord{\left/ {\vphantom {{-c} 
L}}\right.\kern-\nulldelimiterspace}\!\lower0.7ex\hbox{$L$}}}
\thinspace \thinspace \thinspace 
\thinspace \thinspace \thinspace \thinspace \thinspace \thinspace \thinspace (2)
\]

\noindent
where $L$ is the diffusion length and $b$ is an efficiency term to allow for the 
input flux to the diffusion region. We have applied this model iteratively across the neutral 
region in steps of 1 micron to simulate the situation in the MIRI detectors, 
where the absorption can extend well into this region. 

\begin{figure}[htbp]
\centerline{\includegraphics[width=6.0in,height=4.0in]{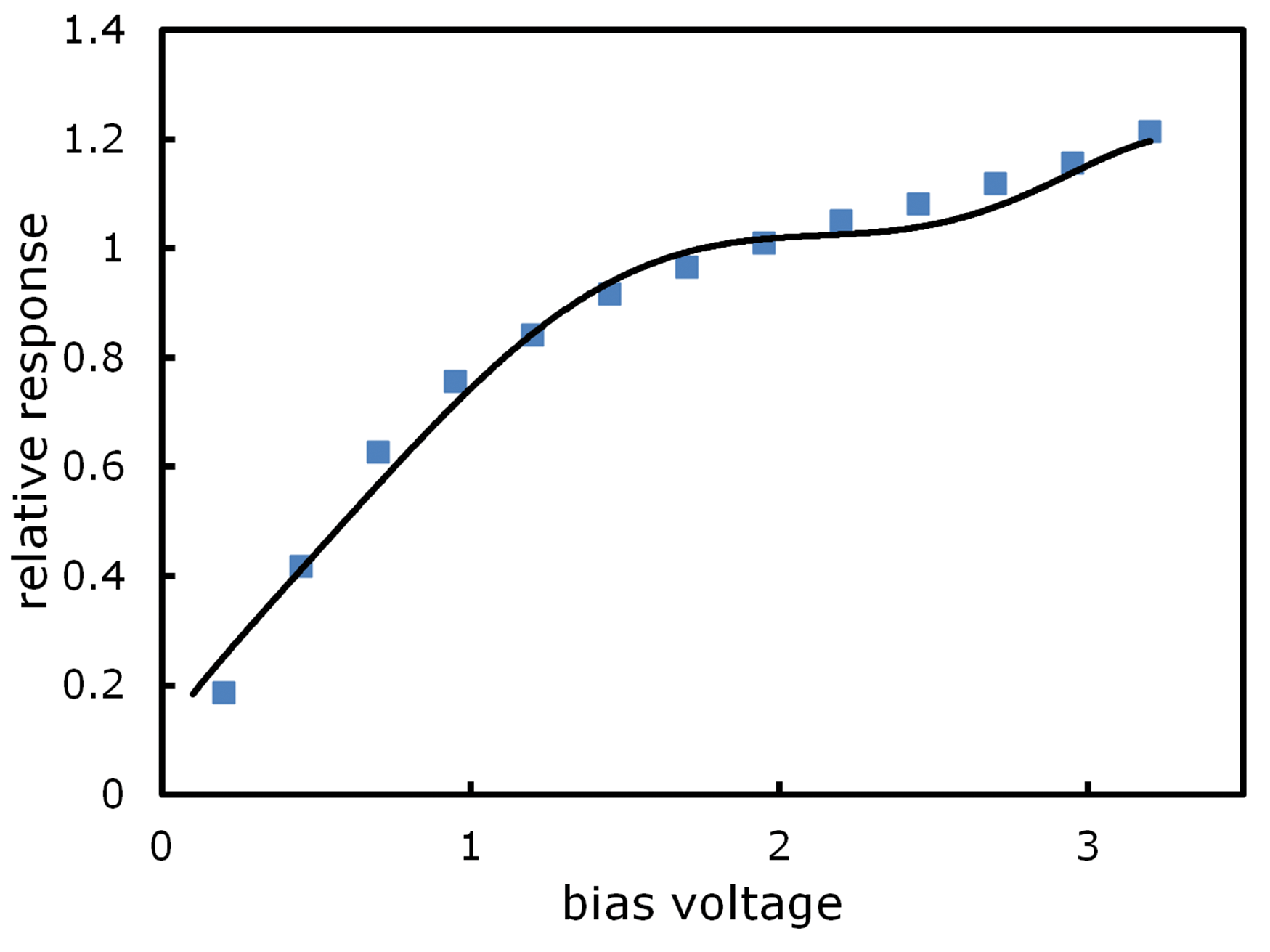}}
\label{fig9}
\caption{Model responsivity vs. bias (line) compared with measurements (points) for the baseline detector material.} 
\label{fig:fig9}
\end{figure}

\begin{figure}[htbp]
\centerline{\includegraphics[width=6.0in,height=4.0in]{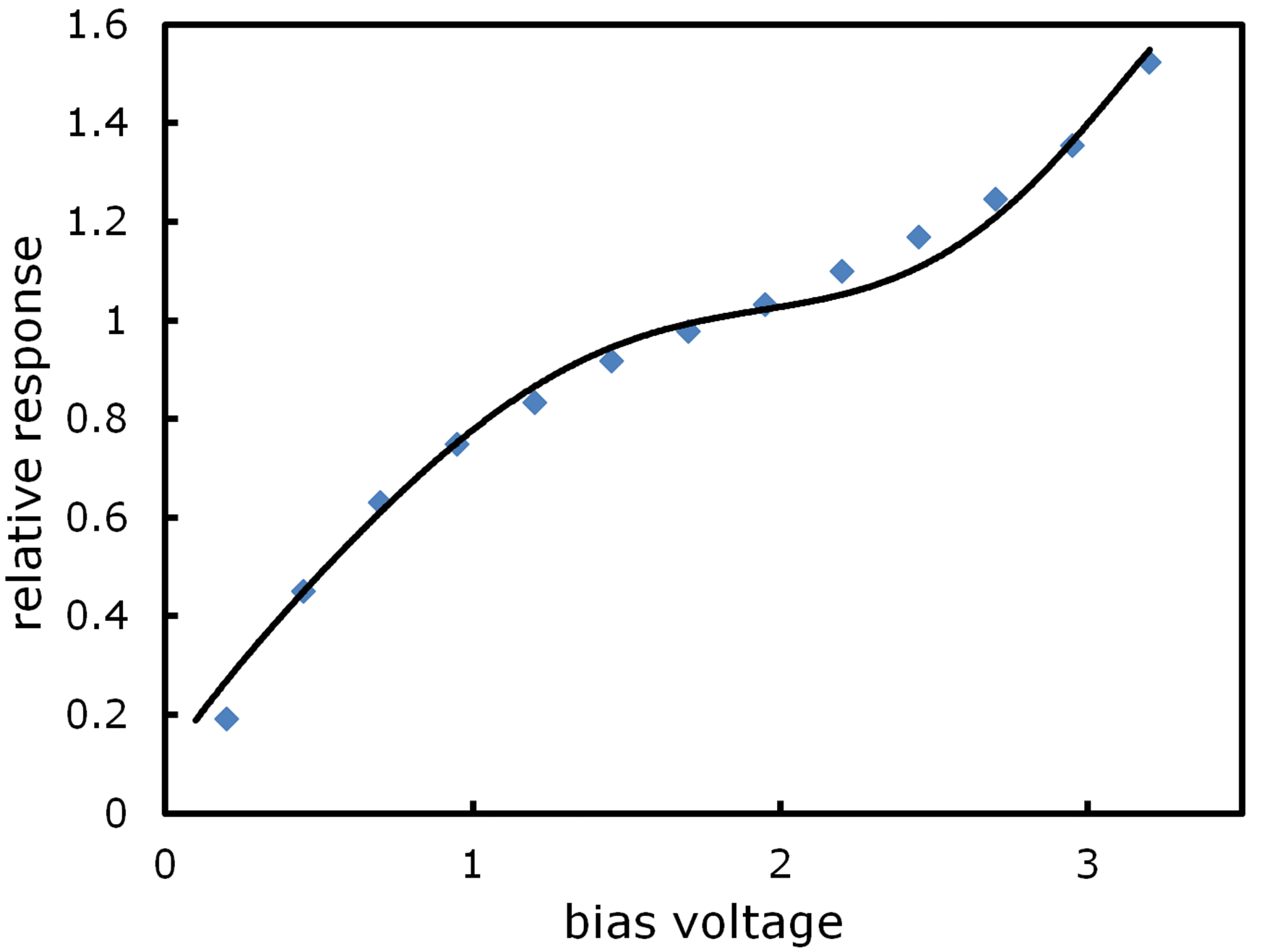}}
\label{fig10}
\caption{Model responsivity vs. bias (line) compared with measurements (points) for the contingency detector material.} 
\label{fig:fig10}
\end{figure}

Figure \ref{fig:fig9} shows the results from a model for the baseline array 
where the impurity concentration, characteristic field strength,
and diffusion length were varied to match the voltage dependence of the response. The parameters 
for this model are $E_{{C}} =$ 8700 V/cm, $N_{{A}} =$ 1.5 
X 10$^{\mathrm{12}}$ cm$^{\mathrm{-3}}$, and $L$ $=$ 2.5 $\mu $m. The 
absorption cross section was set to 5 X 10$^{\mathrm{-16}}$ cm$^{\mathrm{-2}}$ \citep{geist1989}. 
The derived minority impurity 
concentration is consistent with the spreading resistance measurement \citep{love2005}. The 
corresponding value for the recombination time is about 2 X 
10$^{\mathrm{-7}}$ s, assuming a mobility of 3 X 10$^{\mathrm{3}}$ 
cm$^{\mathrm{2}}$ V$^{\mathrm{-1}}$ s$^{\mathrm{-1}}$ from the I. F. Ioffe 
compilation \citep{ioffe2014}. The model indicates that the avalanche gain at the selected bias 
voltage of 2.2V is only a few \%. A second model was produced for the contingency array data, as shown in 
Figure \ref{fig:fig10}. The parameters are $E_{{C}} \quad =$ 8900 V/cm, 
$N_{{A}} \quad =$ 2.3 X 10$^{\mathrm{12}}$ cm$^{\mathrm{-3}}$, and $L$ $=$ 
2 $\mu $m. Except for the minority impurity concentration, they are in good 
agreement with the parameters for the baseline array, which supports the 
validity of the model. Unfortunately, it appears likely that the minority 
impurities are not as well controlled for this material. The avalanche
gain at a bias of 2.2V is indicated to be about 10\%.

Although both models provide reasonably good fits to the observed trend of 
responsivity vs. bias voltage, neither is perfect. An obvious cause of 
discrepancies is that the models assume a single value of the minority 
impurity concentration throughout the IR-active layer. In practice, it is 
likely that the impurity concentration has gradients. Even more 
fundamentally, a 10 $\mu $m thick layer under a pixel will have only about 
10,000 impurity atoms; that is, the rms statistical fluctuations will be at 
the 1{\%} level, sufficiently large to 
influence the responsivity behavior. The changes in the responsivity of a 
given pixel due to just the statistical fluctuations in
the minority impurity concentration under that pixel are large enough to contribute significantly
to the rms pixel-to-pixel responsitivity variations. 

\subsection{Nonlinearity}

Although nonlinearity in near-infrared photodiode arrays is dominated by 
changes in the pixel capacitance, the small fraction of the node capacitance due to the
detectors in the MIRI arrays demonstrates that this mechanism is unimportant for them. 
To be specific, the capacitance at the input to the unit cell
buffer amplifier is 28.5 fF \citep{mcmurtry2005}, the detector is a simple plane-parallel 1.1 fF capacitor,
and the bump bonds may contribute an additional $\sim$ 4 fF to the node capacitance \citep{moore2005}. The
nominal capacitance at the integrating node is therefore 33.6 fF, of which only about 3\% is contributed by
the detector.  

The array amplifiers are linear to within about 1\%, as measured by injecting
signals into the integrating node to calibrate the gain as a function of signal strength. 
The dominant cause of nonlinearity is the reduction in detector responsivity
as signal is accumulated on the amplifier integrating node. When 
the detector bias is reset, the voltage across it will be set accurately
to V$_{bias}$ (see Figure 3) because its effective resistance of $\sim$ $5 \times 10^{19} \Omega$
(estimated from dark current measurements)
is far larger than the ``on" resistance of the reset FET. Once the reset FET
is turned off, as charge accumulates on the amplifer input node capacitance it
reduces the net bias voltage on the detector. The result is that the responsivity
declines as shown in Figure \ref{fig:fig4}. 

\begin{figure}[htbp]
\centerline{\includegraphics[width=4.5in,height=4.0in]{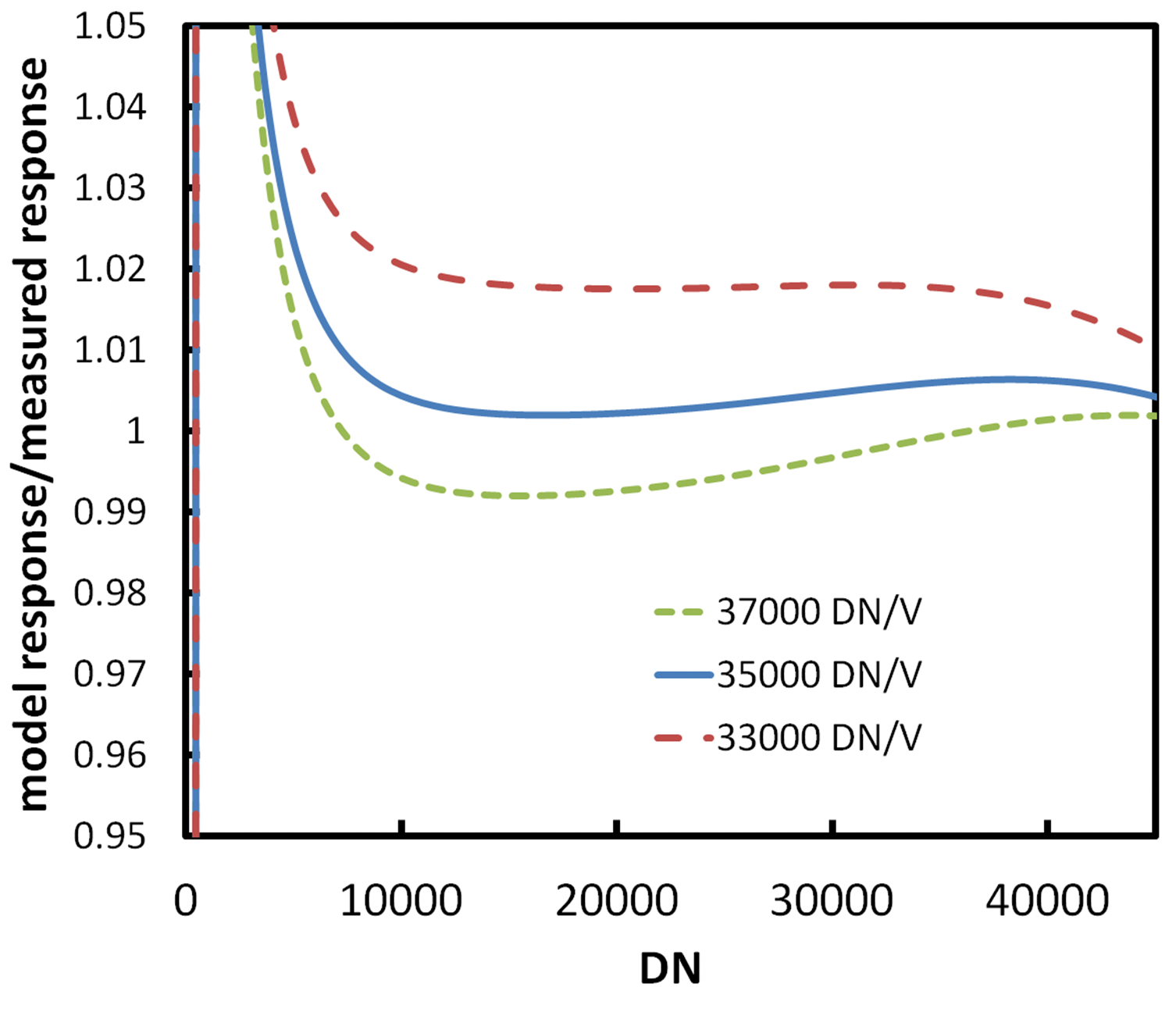}}
\label{fig6a}
\caption{Departure from linearity for different assumptions about the system gain. A zero point of 6000 DN has been subtracted from the integration ramp before taking the ratio relative to the predicted purely linear behavior.} 
\label{fig:fig11}
\end{figure}

A model based on this behavior is able
to reproduce the array nonlinearity accurately. The performance is
illustrated in Figure \ref{fig:fig11}, which shows the ratio of a model signal  assuming 
perfect linear response to the measured signal corrected for
nonlinearity. In computing
this ratio, the only free parameters are the charge accumulation rate (the slope of
the integration ramps) and the gain of the detector/amplifier system (that is the
DN of output per volt change at the integrating node, which determines the relation of the DN of the signal to
the position on the responsivity curve in Figure 4). As shown in Figure 11, the best
fit to the shape of the nonlinearity (that is where the residuals relative to a ratio of 1 are
smallest) is for a gain of about 35000 DN/V. 
The voltage gain of the system has been measured directly to be 38300 DN/V, determined by injecting a signal into
integrating node and measuring the output (Paper VIII). That is, the agreement between the value
deduced indirectly through the model for nonlinearity and the measured one is excellent.  
Combined with the node capacitance,
discussed above, the pixel gain is 5.5 e/DN, with an uncertainty of about 10\%.

As charge is collected, eventually the detector bias
will fall low enough that the depletion depth of the IR-active layer will be
reduced. According to the detector model described in Section 3.1, 
the baseline array is no longer fully depleted at biases of 1.5V and lower, that is 
after collection of a total signal of about 27,000 DN or 150,000 electrons. However, the
loss of depleted depth is fairly slow; at 1V, about 75\% of the IR-active region is
still depleted. A bias of one volt corresponds to an integrated signal of 46,000 DN or 250,000 electrons; it is a useful
limit for "full well" and in fact the MIRI warm electronics saturates not far above it.  Nonetheless,
there are observable consequences from the loss of full depletion: 1.) for large signals, the detective 
quantum efficiency of the detectors will be reduced; and 2.) because of the different absorption profiles
in the IR-active layer for different wavelengths, the nonlinearity will have a wavelength dependence.

\subsection{Imaging properties}

\subsubsection{Latent images}

Unfortunately, there appears to be little previous analysis of potential causes of latent images in these arrays, so we will
make some suggestions without coming to firm conclusions. 
Low levels of trapped charge could result from impurities in the blocking layer, which might cause
weak latent images at all levels of illumination. Another possibility is 
dielectric relaxation - the slow adjustment of the electrical equilibrium 
of some portion of the detector, basically because of its long RC time constant. 
Stapelbroek (private communication) has calculated that the time constant for
this process in the undepleted region of the detectors is roughly appropriate to 
lead to the more rapidly decaying latent images. \citet{smith2008} propose that latent images in photodiodes
can arise because the depleted region at the junction shrinks as the detector is debiased
due to accumulation of signal, allowing trapping of photoelectrons. 
Once the bias is re-established, the traps lie within the depletion region; as the trapped electrons
are slowly released, they appear as a signal. Although this argument cannot be applied directly
to Si:As IBC detectors, a similar situation may exist if sufficient signal is gathered to
produce a significant undepleted region (from the above section, a signal of order
200,000 electrons would suffice). A number of these processes could operate together. 
Thus, why the latent image behavior might be complex is not difficult to understand, but
a systematic set of measurements as a function of detector temperature, photon flux, and
perhaps wavelength is needed to obtain sufficient information to constrain the possibilities. 

\subsubsection{Image 'halos' at short wavelengths}

Observations with IRAC (in spectral bands at 5.8 and 8 $\mu $m) showed that 
a significant fraction of light was being detected in what was described as 
a halo around the core image (Fazio et al. 2004; Hora et al. 2004b; Pipher et al. 2004; Figure 8). This 
behavior was also seen in laboratory measurements (C. McMurtry, private 
communication), where the phenomenon appears to be more accurately described 
as light scattered along the rows and columns of the array (e.g., a cross 
geometry centered on the image core). The behavior is specific to operation 
of this type of detector array at wavelengths between 5 and 10 $\mu $m (and 
presumably at shorter ones, but the devices are seldom used short of 5 $\mu $m). 
The clearest demonstration that this behavior is absent at longer 
wavelengths is the excellent agreement between the point and extended source 
calibrations at 24 $\mu $m for MIPS \citep{cohen2009}. Figures 7 and 8 
demonstrate that the MIRI arrays have very similar imaging performance to those 
in IRAC and MIPS, i.e., clean images at wavelengths longer than 10 $\mu$m, but 
cross-like artifacts at shorter wavelengths.

An explanation of this behavior is based on the 
geometry illustrated in Figure 2. The detectors are fabricated on a 
silicon substrate $\sim$ 500 $\mu $m thick that is transparent in the mid-IR. Light enters this 
substrate at the top of the figure. Even with AR coatings, reflectivities in the 5 - 8$\mu$m range 
are of order 8{\%} \citep{hora2004b, love2005}. The mid-IR photons are absorbed in the 
IR-active layer. Between it and the bulk of the substrate there is a thin 
transparent buried contact, which can also reflect some of the light; 
nominal reflectivities can be up to 20{\%} \citep{petroff1985}, or 
25{\%} \citep{szmulowicz1988}. The reflection losses are also apparent
from Figure 5, since the responsive quantum efficiency never gets above
about 80\% even at wavelengths where the absorption in the IR-active layer 
should be complete and the AR coating highly efficient, e.g., $\sim$ 16 $\mu$m. 

For the $7 \times 10^{-17}$ 
cm$^{\mathrm{-3}}$ arsenic concentration in the IR-active layer of the MIRI baseline detectors, the 
imaginary component of the index of refraction is \citep{woods2011}:

\[
\thinspace k=5.69\times {10}^{-3}\left( \frac{\lambda }{7\thinspace \mu 
\mathrm{m}} \right)^{3}\thinspace \thinspace \thinspace \thinspace 
\thinspace \thinspace \thinspace \thinspace \thinspace \thinspace \thinspace 
(3)
\]

\noindent
Here $\lambda $ is the wavelength. The corresponding linear absorption 
coefficient is:
\[
\thinspace \alpha =102\thinspace \left( \frac{\lambda }{7\thinspace \mu 
\mathrm{m}} \right)^{2}{\mathrm{cm}}^{-1}\thinspace \thinspace \thinspace 
\thinspace \thinspace \thinspace \thinspace \thinspace \thinspace \thinspace 
(4)
\]
(This formulation is not applicable at the longer wavelengths, e.g., 
\textgreater 15 $\mu $m.). For the 35 $\mu $m active layer thickness of 
the MIRI baseline detectors, the absorption at normal incidence is 17{\%}, 65{\%}, and 95{\%} at 
wavelengths of 5, 12, and 20 $\mu $m for a single pass through the IR-active 
layer. Most of the light not absorbed will be reflected by the pixel contact 
to make a second pass, so the total absorption will be about 28{\%}, 83{\%}, 
and 100{\%} respectively at these three wavelengths. At short wavelengths, a 
significant amount of light survives after encountering the active layer. To 
understand the cross artifact, we need to understand the fate of this light.

Simple arguments show that the cross artifacts must be associated with the interpixel gaps (see Figure  \ref{fig:fig2}). 
Optically, these gaps will act as narrow slits, with widths comparable to 
the wavelength of the light. Therefore, we cannot use a geometric approach 
to understand their effects, since they will spread the light by 
diffraction. Traditionally, diffraction is treated in terms of the effects 
on the light transmitted through a slit. However, Babinet's Principle (\citet{hecht2001}, Section 10.3.11) states that the Fraunhofer diffraction patterns of 
complementary masks are identical, that is the equivalent diffraction 
pattern will be obtained if the slit is replaced by an opaque occulter with 
free transmission in the surround. This situation obtains for the reflected 
light around an interpixel gap in the detector arrays (with the addition of a fold of the beam through the reflection). 
We can estimate the angular dependence of the diffracted light in terms of
a mm-wave analog from
\citet{lee2009}\footnote{Other studies of the effects of narrow slits on optical wavelength
light tend to use slits with thicknesses larger than the photon wavelength and the emergent light
angular distribution is strongly affected by waveguide effects.}, who illuminated slits 10 $\mu$m thick 
with light of wavelength between 6000 and 150 $\mu$m, comparable to the ratio of wavelength
to slit thickness of about 50 for the contacts on a Si:As IBC array. \citet{lee2009} find transmissions approximating the
geometric thin slit case for frequencies up to a cutoff of c/2a where c is the speed of light and a is the slit width.
In the case of the Si:As IBC arrays, the physical interpixel gaps (slit widths) are effectively magnified by the refractive
index of silicon, so they can be considered to be 27 $\mu$m (IRAC) and 10 $\mu$m (MIRI), and fall into
this pseudo-geometric regime. 
Thus, far from the gaps the illumination pattern should approximate classical Fraunhofer diffraction. 

Therefore, a single-slit diffraction pattern will be imposed on the 
reflected beam equivalent to that of a slit with the geometry of the 
interpixel gap. Some of the diffracted light will be reflected off the buried contact
and the input surface of the detector to be detected removed from the pixel where it entered the array. 
The spreading will be perpendicular to the slits, which 
explains why the artifacts are along rows and columns of the array. 
However, simple diffraction and reflection do not account for the extent of the cross 
artifact. A more detailed model shows that some of the light is diffracted 
beyond the angle for trapping by total internal reflection in the detector. 
In addition, some of the light escapes the detector wafer through the 
interpixel gaps into the space between the detector and the readout circuit
 (see Figure 2). 
This region acts like a lossy integrating cavity, and some of the light can 
re-enter the detector wafer through other interpixel gaps. Again, this light 
may be diffracted to angles where the photons are trapped by 
total internal reflection. Because the entire wafer is inefficient at 
absorbing photons in the 5 $\mu $m region, those that are trapped can travel 
large distances before they are absorbed and produce signals. This behavior
can account qualitatively for the cross image artifacts; a more detailed discussion
will be published elsewhere.

\section{Conclusion}

The Si:As IBC detector arrays for MIRI represent the next step in the development
of detectors for the 5 - 28 $\mu$m spectral range, building on the heritage from the
detectors used in the three {\it Spitzer} instruments, WISE, MSX, and Akari. They
have a larger 1024 X 1024 pixel format, and show good quantum efficiency and low dark
current. We have modeled the detector behavior to understand the avalanche gain
characteristics, the behavior of their nonlinear response, the effect of large
signals on their absorption efficiency, and the extended images found near 5 $\mu$m. 
This work will help optimize the operating parameters for the detectors and provides
inputs to the data pipeline development for MIRI.

\section{Acknowledgements}
The work presented is the effort of the entire MIRI team and the enthusiasm within the MIRI partnership is a significant factor in its success. MIRI draws on the scientific and technical expertise many organizations, as summarized in Papers I and II. 
A portion of this work was carried out at the Jet Propulsion Laboratory, California Institute of Technology, under a contract with the National Aeronautics and Space Administration.

In addition, we thank Scott Knight, Roger Linfield, and "Dutch" Stapelbroek for helpful discussions. Thanks also to 
Analyn Schneider and Johnny Melendez of JPL who put in untold numbers of hours obtaining much of the data taking
summarized here. Part of the research described in this paper was carried out at the Jet Propulsion Laboratory, California Institute of Technology, under a contract with the National Aeronautics and Space Administration. This work 
was also supported in part by NASA grant NNX13AD82G to G. H. Rieke.

\clearpage

\begin{deluxetable}{ccc}
\tabletypesize{\footnotesize}
\tablecolumns{3}
\tablewidth{0pt}
\tablecaption{Detector Performance Parameters}
\tablehead{\colhead{Parameter}                   &
           \colhead{baseline array}          &
           \colhead{contingency array}         }
       
\startdata
format	&	1024 x 1024	&	1024 x 1024	\\
pixel size	&	25 $\mu$m	&	25 $\mu$m 	\\
IR-active layer thickness  &   35 $\mu$m   &   30 $\mu$m   \\
IR layer As doping   &  $7 \times 10^{17}$ cm$^{-3}$  &   $ 5 \times 10^{17}$  cm$^{-3}$   \\   
read noise*	&	14 e$^-$	&	14 e$^-$	\\
dark current  &        0.2 e$^-$/s     &       0.07 e$^-$/s         \\
quantum efficiency**   &   $\ge$ 60\%   &   $\ge$ 50\%  \\
nominal detector bias***  &   2.2V          &       2.2V         \\
well capacity   &     $\sim$ 250,000 e$^-$      &      $\sim$ 250,000 e$^-$  \\

\enddata
\tablenotetext{*}{Fowler-eight sampling, used for comparison purposes; the readout is normally operated in a sample-up-the-ramp mode.}
\tablenotetext{**}{At peak wavelength}
\tablenotetext{***}{Consisting of 2 V applied directly plus $\sim$ 0.2 V from clocking signal feedthrough}
\end{deluxetable}

\clearpage

\clearpage





\clearpage

\clearpage

\clearpage

\clearpage

\clearpage

\clearpage

\clearpage

\clearpage

\clearpage

\clearpage

\end{document}